\documentclass{llncs}
\usepackage{graphics,graphicx}
\usepackage{amsmath}
\usepackage{wrapfig}
\usepackage{subcaption} 
\usepackage{xcolor}

\usepackage{amssymb,bm,amsmath,amsfonts,mathrsfs}
\usepackage[ruled, vlined]{algorithm2e}

\def\bdf{\fontseries{b}\selectfont}
\begin{document}

\title{Globally Optimal Surface Segmentation using Deep Learning with Learnable Smoothness Priors}

\author{Leixin Zhou, Xiaodong Wu}
\institute{Department of Electrical and Computer Engineering \\The University of Iowa, Iowa City, USA \\
\email{leixin-zhou@uiowa.edu, xiaodong-wu@uiowa.edu}
}

\maketitle
\begin{abstract}
Automated surface segmentation is important and challenging in many medical image analysis applications. Recent deep learning based methods have been developed for various object segmentation  tasks. Most of them are a classification based approach, e.g. U-net, which predicts the probability of being target object or background for each voxel. One problem of those methods is lacking of topology guarantee for segmented objects, and usually post processing is needed  to infer the boundary surface of the object. In this paper, a  novel model based on convolutional neural network (CNN) followed by a learnable surface smoothing block 
is proposed to tackle the surface segmentation problem with end-to-end training. To the best of our knowledge, this is the first study to learn smoothness priors end-to-end with CNN for direct surface segmentation with global optimality. Experiments carried out on Spectral Domain Optical Coherence Tomography (SD-OCT) retinal layer segmentation and Intravascular Ultrasound (IVUS) vessel wall segmentation demonstrated very promising results.
\end{abstract}

\section{Introduction}
Automated segmentation of objects or equivalently boundary surfaces plays a very import role in quantitative image analysis. In several years, deep learning based method for semantic segmentation has become very popular in computer vision and medical imaging. The fully convolutional networks (FCN) \cite{long2015fully}
, and then U-net
\cite{ronneberger2015u}
  for medical image segmentation have been proposed. All these methods model the segmentation problem as a pixel-wise or voxel-wise classification problem. Popular loss functions for training these networks include cross entropy (CE), weighted cross entropy (WCE) for imbalanced classes, and dice similarity coefficient (DSC).
As there is no explicit constraint among  the labeling of different pixels,  usually post processing, such as morphological operation, is required to get reasonable prediction.
\par
It is well known that prior knowledge, like shape priors, can help significantly improve segmentation performance, especially in medical scenario the target objects often lack strong boundaries, and/or present similar intensity profiles with multiple crowed nearby tissues. Some studies have investigated along this line. For example, Milletari {\em et al.}~\cite{milletari2017integrating} proposed to integrate into CNN the statistical shape priors, which were  obtained from principal components analysis (PCA). Although the robustness of segmentation was improved, the mean DSC did not increase. Ravishankar {\em et al.}~\cite{ravishankar2017learning} utilized a shape regularization network to project the preliminary prediction into shape space, and a shape loss was added to make the prediction more fit into the desired shape space. Chen {\em et al.}~\cite{chen2018end} proposed an end-to-end approach with CNN and  Conditional Random Fields (CRFs) for segmentation. However, the inference of CRFs was approximate and not globally optimal.
\par
Inspired by the graph search method \cite{li2006optimal},  Shah \textit{et al}. \cite{shah2018multiple} first modeled the terrain-like surfaces segmentation as direct surface identification using regression. The network consists of an FCN followed by fully-connected layers. 
The network   is  very light weighted and no post processing is required.  Surprisingly the results are very promising~\cite{shah2018multiple}. It is well known that U-net outperforms FCN because U-net has an additional expansive path such that features of high resolution can be learned and then better prediction accuracy can be achieved. To improve segmentation accuracy further, it is natural to think to replace the FCN with a U-net. However, it is not reasonable to concatenate a U-net with fully-connected layers, as the invariance of feature maps in the original resolution is supposed to be much less than that in the low resolution, such that there would be much more chance that the fully-connected layers heavily overfit to the training data. Another drawback of Shah \textit{et al}'s work  ~\cite{shah2018multiple} is that the surface smoothness is {\em implicitly} learned within the whole network (mainly within fully-connected layers) as a black box. It is hard to decode after training. 
\par
To resolve problems mentioned above, we propose to explicitly model the surface segmentation problem as a quadratic programming, which can be solved with guaranteed global optimality. The whole network can be trained end-to-end. Our contributions are in three folds: 1) The first time to parameterize the output of the U-net as Gaussians (the mean represents surface position prediction from the U-net and the standard deviation encodes the prediction confidence), which converts the description from the discrete to the continuous space; 2) The parameter that controls smoothness can be learned and has a clear meaning; 3) The method works in the continuous space and enables sub-pixel segmentation.

\section{Method}
\begin{wrapfigure}{r}{0.3\linewidth}
	\vspace{-2.cm}
	\centering
	\includegraphics[width=0.3\textwidth]{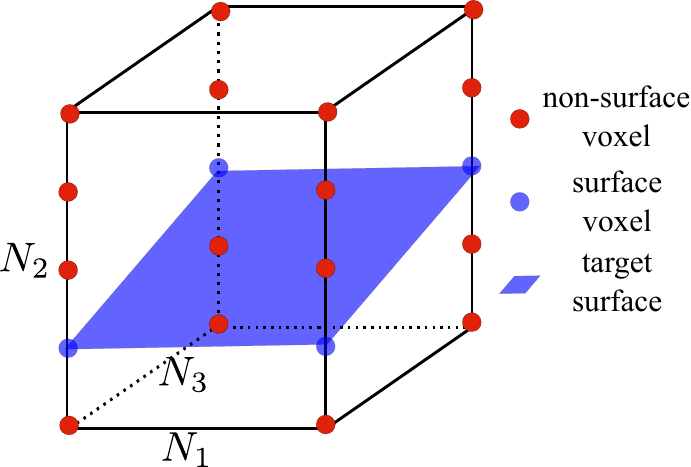}	
	\vspace{-.8cm}
	\caption{Surface segmentation definition.}
	\label{fig:surf_def}
	\vspace{-0.8cm}
\end{wrapfigure}
We first define the surface segmentation problem.
A $3$-D image can be viewed as a $3$-D tensor $\mathcal{I}$. A terrain-like {\em surface} in $\mathcal{I}$ is oriented and shown in Fig.~\ref{fig:surf_def}. Let $N_1$, $N_2$ and $N_3$ denote the image sizes in three dimensions, respectively. Let all column index be a set $\Omega=\{(1,1),(1,2),\cdots,(N_1,N_3)\}$. The surface $\mathbf{x}$ is defined by $x_i \in [1, N_2]$, $\forall i \in \Omega$. Thus any feasible surface in $\mathcal{I}$ intersects with  each {\em column} exactly once.
Generally in surface segmentation~\cite{li2006optimal}, the problem is formulated as minimizing the energy function $E(\mathbf{x}): \mathbb{R}^{N_1\times N_3}\rightarrow \mathbb{R}$
\begin{align}\label{Defn:OverallEnergyFunction}
E(\mathbf{x}) = E_u(\mathbf{x}) +  E_p(\mathbf{x}),
\end{align}
where the unary term $E_u$ is the energy when considering each {\em column} independently, and the pairwise energy term $E_p$ penalizes discontinuity of surface position among adjacent {\em columns}. The design of $E_u$ and $E_p$ will be detailed in Section~\ref{sec:prop}.
\subsection{Proposed Inference Pipeline}\label{sec:prop}
\begin{figure}
	\vspace{-0.9cm}
	 \captionsetup[subfigure]{aboveskip=-1pt,belowskip=-10pt}
	\centering
	\begin{subfigure}[b]{0.68\textwidth}
		\includegraphics[width=\textwidth]{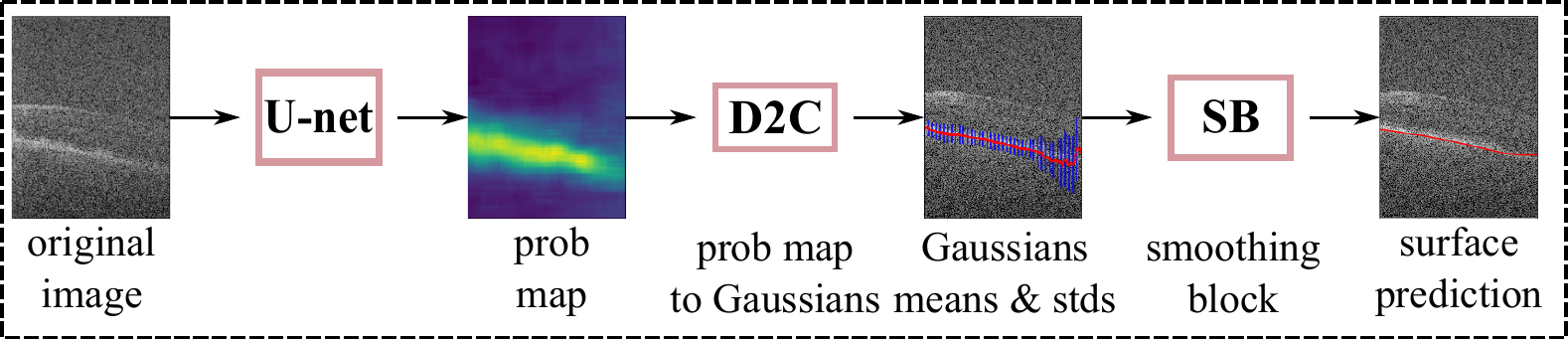}
		\caption{}
		\label{fig:infer}
	\end{subfigure}
	\begin{subfigure}[b]{0.3\textwidth}
		\includegraphics[width=\linewidth]{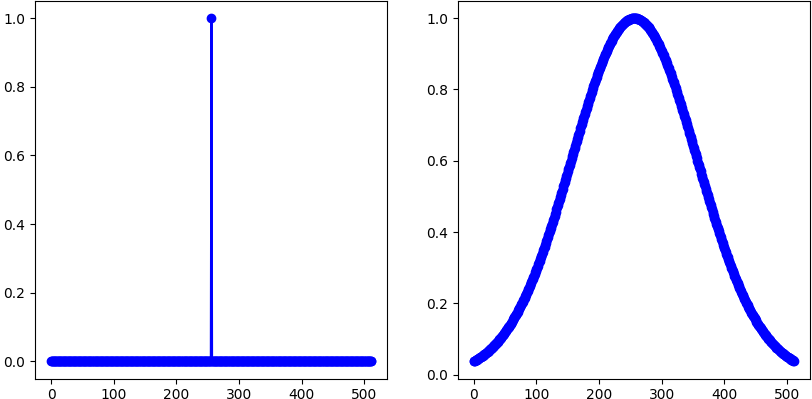}
		\caption {}
		\label{fig:loss_comp}
	\end{subfigure}
	\caption {(a) Inference pipeline of the proposed method. (b)  Target surface position (left) and its relaxed surface possibility map (right) for one column.}
	\vspace{-0.9cm}
\end{figure}

{\em One should note that the proposed method can be applied in both $2$-D and $3$-D.} For the purpose of proof-of-concept and clear explanation, the detailed description of the proposed method and all experiments are done in $2$-D. The inference pipeline of the proposed method is demonstrated in Fig.~\ref{fig:infer}.  The trained U-net takes in the original image $\mathcal{I} \in \mathbb{R}^{N_2\times N_1}$, and outputs the discrete probability map $P \in \mathbb{R}^{N_2\times N_1}$. Ideally, for each image column, the probability for the target surface position is high, and it  is gradually reduced on the positions away from the one on the surface, as demonstrated in Fig.~\ref{fig:loss_comp}. We thus propose a  block to convert the discrete probability map $P$ to a Gaussian parameterization $G \equiv (\mathbf{\gamma}, \mathbf{\sigma}) \in \mathbb{R}^{N_1 \times 2}$, where $\gamma_i$ specifies the mean surface position on each column $i$  and $\sigma_i$ is the corresponding standard deviation. The Gaussian parameterization $G$ is then  fed into the trained smoothing block (SB) , which incorporate the learned surface smoothness priors to infer the optimal target surface. Next, we detail the novel blocks in our deep optimal surface segmentation neural network.
\par \noindent {\bf D2C block}
The D2C block is basically designed to convert the discrete probability map of each column $P_i \in \mathbb{R}^{N_2}$ $\forall i \in \{1,2,\cdots,N_1\}$, which is output from U-net, to a continuous representation $G_i \equiv (\gamma_i, \sigma_i) \in \mathbb{R}^2$ (Gaussian is utilized in our design). This enables optimizing directly on the surface position and sub-pixel accuracy prediction. 
The proposed conversion is realized by fitting a continuous Gaussian function to the discrete probability map $P_i$, which can be thought as discrete samples of a continuous Gaussian probability density function.
Recall that one dimensional Gaussian function has the formula  $f(j) = A\exp(\frac{-(j-\gamma)^2}{2\sigma^2}),$
and then $\ln (f(j)) = \ln (A) + \frac{-(j-\gamma)^2}{2\sigma^2} = \ln (A) - \frac{\gamma ^2}{2\sigma^2} + \frac{2\gamma j}{2\sigma^2} - \frac{j^2}{2\sigma^2} = a + bj + cj^2,$
where $a= \ln (A) - \frac{\gamma ^2}{2\sigma^2}$, $b=\frac{\gamma}{\sigma^2}$, and $c=- \frac{1}{2\sigma^2}$.
In our setting, for each column, we have $N_2$ samples of $(j, f(j))$. We can define an error function  \cite{guo2011simple}, namely $\varepsilon = \sum_{j=1}^{N_2} f(j)^2(\ln (f(j)) - (a+bj+cj^2))^2.$
Then minimizing the weighted mean square error (MSE) $\varepsilon$, one can get the estimates of $a, b, c$ by solving a set of three linear equations and then $A, \gamma, \sigma$. The problem is very similar to least square problem~\cite{guo2011simple}. 
In our implementation, a linear transform is utilized to normalize the probability map for each column to the range [0,1], then we can ignore the magnitude $A$. As the computation of each column is independent, it is straightforward to be extended to $3$-D.
\par \noindent {\bf Smoothing Block (SB)}
To integrate the surface segmentation model (Eq.~\ref{Defn:OverallEnergyFunction}) with smoothness priors, we define the energy function $E(\mathbf{x}): \mathbb{R}^{N_1}\rightarrow \mathbb{R}$ as
\vspace{-0.25cm}
\begin{align}\label{Defn:EnergyFunction}
E(\mathbf{x}) &= \sum_{i\in\Omega} \psi_u(x_i) + \sum_{\substack{i,j\in\Omega, (i,j)\in\mathcal{N} }} \psi_p(x_i,x_j),
\end{align}
\vspace{-0.1cm}
where $\psi_u(x_i)$ and $\psi_p(x_i,x_j)$ are defined as $\psi_u(x_i) =  \frac{(x_i - \gamma_i)^2}{2\sigma_i^2},~~\psi_p(x_i,x_j) = w_{comp} \cdot (x_i - x_j)^2,$
and $\mathcal{N}$ is the set of neighbor columns.
\begin{wrapfigure}{r}{0.3\linewidth}
	\centering
	\vspace{-.8cm}
	\includegraphics[width=0.3\textwidth]{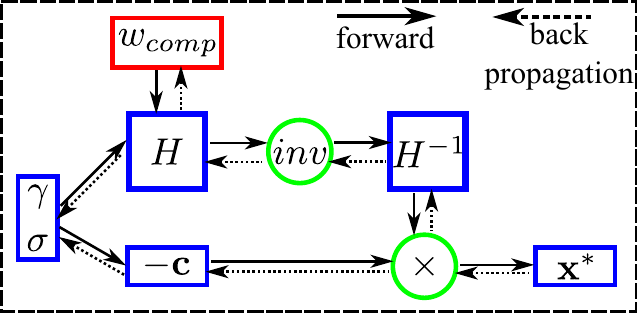}	
	\caption{The proposed SB architecture. Only one parameter $w_{comp}$ needs to be trained.}
	\label{fig:sb}
	\vspace{-0.5cm}
\end{wrapfigure}
For simplicity, the nearest neighbor pairs, i.e. $\mathcal{N} = \{(1,2), (2,3),\cdots,(N_1-1,N_1)\}$,  are considered as the set of neighbor columns.
The whole energy in Eq.~\ref{Defn:EnergyFunction} can be reformulated as the standard quadratic form
$E(\mathbf{x}) = \frac{1}{2} \mathbf{x}^{T}H\mathbf{x} + \mathbf{c}^{T}\mathbf{x} + \text{const}.$
%
It can be proved that the Hessian matrix $H$ is positive definite by using Gershgorin circle theorem and then the energy function is convex. The gradient is
$\nabla = H\mathbf{x} + \mathbf{c}.$
Let the gradient to be zero, we have the global optimal solution
$\mathbf{x}^{*} = -H^{-1}\mathbf{c}.$ Another advantage of the proposed energy formulation is the optimal solution can be calculated in one step, i.e. we do not need to make use of  a recurrent neural network (RNN)  to implement SB. It is also straightforward to implement the smoothing block in 3D.

\subsection{Training Strategy}
\subsubsection{U-net Pre-training}
\begin{figure}[htb!]
	\centering
	\includegraphics[width=0.7\textwidth]{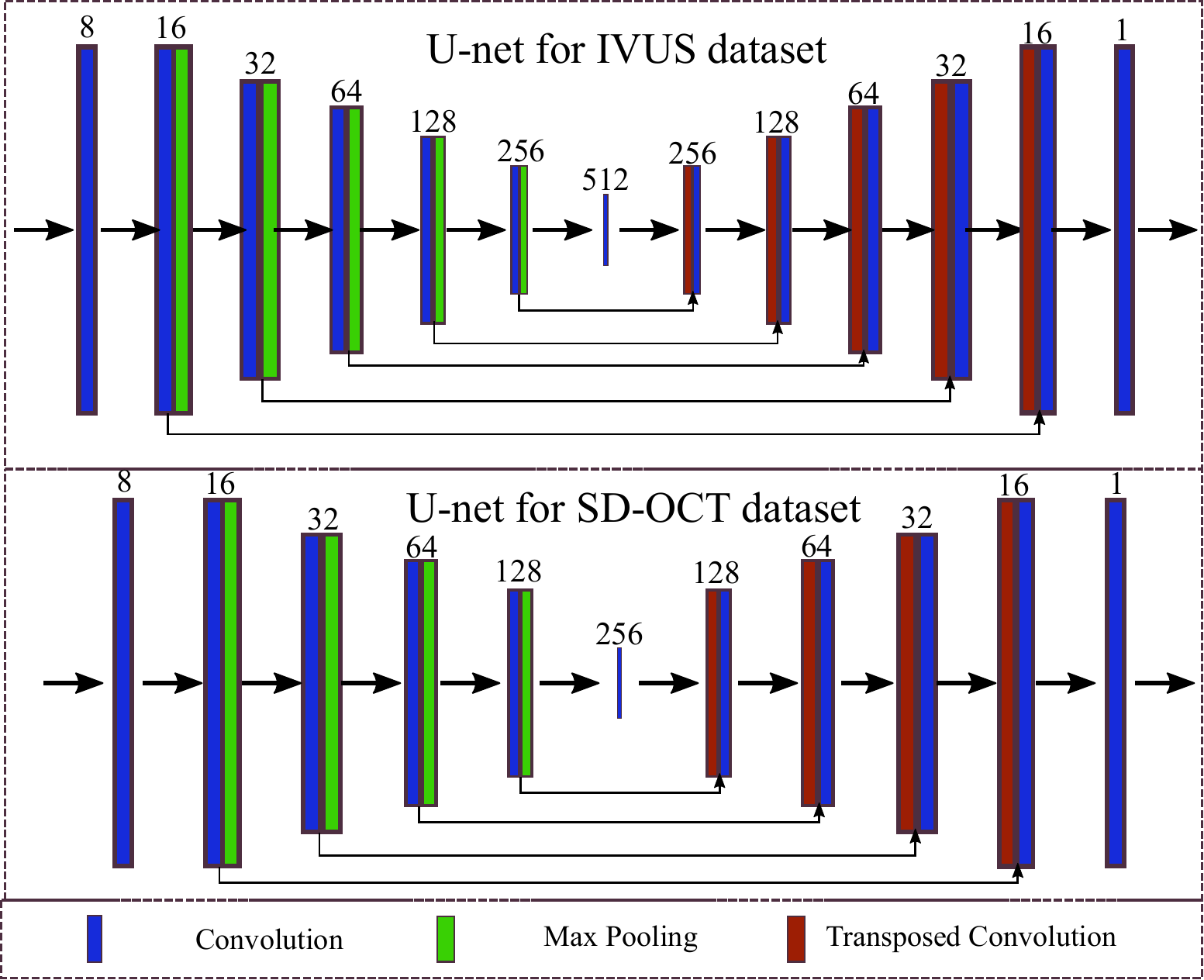}
	\caption{U-net architectures. For convolution layers, the number indicates dimensions of feature maps, and $3 \times 3$ convolution kernels are utilized except the last convolution layer, where $1 \times 1$ kernels are chosen; for max pooling and transposed convolution layers, $2 \times 2$ kernels and strides 2 are utilized. A Relu layer is utilized following each convolution layer except the final $1 \times 1$ convolution layer. For all skipping connections, the features are merged by adding.}
	\label{fig:arch}
\end{figure}

A common U-net architectures is utilized to generate the discrete probability map for the input image ($N_2\times N_1$). The detailed architecture is demonstrated in Fig.~\ref{fig:arch}. In the proposed method, the softmax layer works on each {\em column}, not on each pixel. The rational is that we assume the target surface intersects with each column by exactly once, and so the probabilities are normalized within each column.
Also we assume the U-net should output a Gaussian shaped probability map for each column, which mimics the Bayesian learning  for each column 
and shares merits with knowledge distillation~\cite{hinton15} and distillation defense~\cite{papernot2016distillation}.
To encourage the U-net outputs reasonable probability maps, the Kullback–Leibler divergence (KLD) loss is utilized for the U-net pre-training. 
KLD is a measure of how one probability distribution is different from a reference probability distribution. It is equivalent to the Cross Entropy when the reference is a Kronecker delta function. We propose to relax the delta function to a Gaussian distribution such that the proposed D2C block can work properly. 
We pick $\sigma$ to be around 0.1 times of the column length and our method is insensitive to $\sigma$. Then we will have one Gaussian distribution ground truth for each column, denoted as $T^{\sigma}_i \in \mathbb{R}^{N_2}$. One illustration is demonstrated in Fig.~\ref{fig:loss_comp}. We denote the output from the U-net as $P \in \mathbb{R}^{N_2\times N_1}$ and the loss for the pre-training is formulated as $loss_{pre} (P, T^{\sigma})= -\sum_{i\in\Omega} D_{KL}(T^{\sigma}_i || P_i)).$


\subsubsection{Fine Tuning}	
\begin{figure}
	\vspace{-.8cm}
	\centering
	\includegraphics[width=0.9\textwidth]{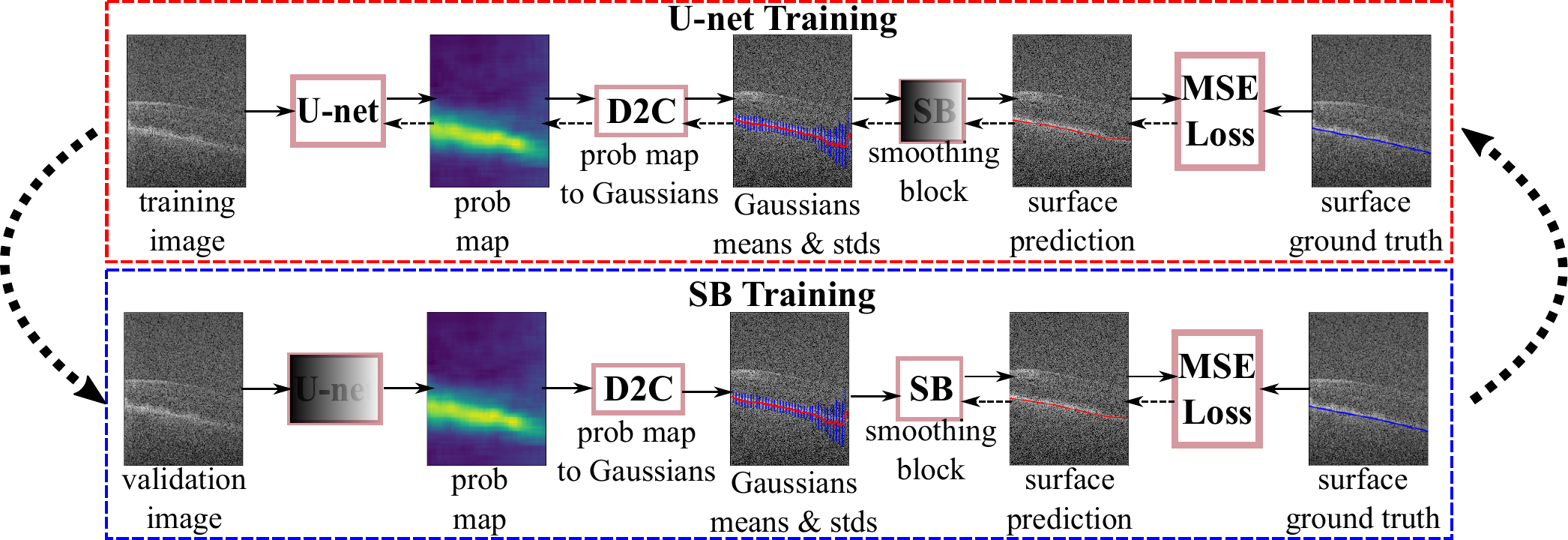}
	\vspace{-0.3cm}
	\caption {Two phases of fine tuning of the proposed network. The fine tuning alternates between training U-net for $EP_{\text{U-net}}$ epochs and training SB for $EP_{\text{SB}}$ epochs. The shaded blocks are kept fixed during the respective training phase.}
	\label{fig:finetune}
	\vspace{-0.7cm}
\end{figure}

During the fine tunning phase, the mean square surface position error (MSE) is utilized as the loss function, which is formulated as $loss_{fine}(\mathbf{x}, \mathbf{t}) = \sum_{i\in\Omega} (x_i - t_i)^2,$
where $\mathbf{t} \in \mathbb{R}^{N_1}$ denotes the ground truth surface positions. The fine tuning of the whole network proceeds in an alternation fashion (Fig.~\ref{fig:finetune}). 
The {\em validation} data is utilized to train the SB, and the training data is for U-net training. As SB only has one parameter-$w_{comp}$ to train, the overfitting chance of it is very low. Also the U-net is not trained on validation data, the learned $w_{comp}$ should be more representative in the wild. Otherwise if fine tuning  U-net+SB simultaneously on the {\em training} data, the learned  $w_{comp}$ is generally  smaller than necessary, as the pre-trained U-net generally has fit the training data well and SB does not play an important role. 
\section{Experiments}
\subsection{SD-OCT Retinal Layer Segmentation}
\begin{wrapfigure}{R}{0.32\linewidth}
		\vspace{-3.3cm}
	\centering
	\includegraphics[width=0.32\textwidth]{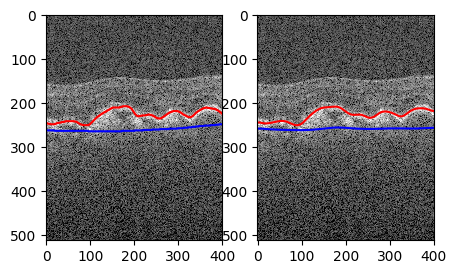}
		\vspace{-.7cm}	
	\caption{Sample SD-OCT data. Ground truth (left) and predictions (right). Red: IRPE; Blue: OBM.}
	\label{fig:oct}
	\vspace{-1.cm}
\end{wrapfigure}
The proposed method was applied to retinal layer segmentation in SD-OCT images, which were obtained from the public dataset \cite{farsiu2014quantitative}.
 Since the manual tracings were only available for a region centered at the fovea, subvolumes of size $400\times 60\times512$ were extracted around the fovea. The dataset was randomly divided into 3 sets: 1) Training set -  263 volumes (79 normal, 184 with age-related macular degeneration (AMD)); 2)Validation set - 59 volumes (18 normal, 41 AMD); 3) Testing set - 58 volumes (18 normal, 40 AMD). The surfaces considered are S2-Inner Aspect of Retinal Pigment Epithelium Drusen Complex (IRPE) and S3-Outer Aspect of Bruch Membrane (OBM) as shown in  Fig. \ref{fig:oct},  which are very challenging to segment. 
\par \noindent {\bf Pre-processing and Augmentation}
The intensity of each slice was normalized to the range [-1, 1]. No additional pre-processing methods were utilized. For the purpose of pre-training the U-Net, the standard deviation of the Gaussian model of the surface position on each column was set   $\sigma=50$. 
 We augmented the training data by applying mirroring along the horizontal dimension, and  2 random translations along the axial dimension on the original image and the mirrored image. 
\par\noindent {\bf Hyperparameters}
All training utilized Adam optimizer. For U-net pre-training, the learning rate was $10^{-4}$, and the training ran for 2000 epochs. For fine-tuning, the learning rate of  U-net ($LR_{\text{U-net}}$) was $10^{-5}$, the learning rate of SB ($LR_{\text{SB}})$ was $10^{-2}$, the total epochs for the two phases were: $EP_{\text{U-net}}=10$, and $EP_{\text{SB}}=10$. The initial smoothness parameter $w_{comp}^{init}$ was set to $10^{-5}$. 
\begin{table}
	\vspace{-0.5cm}
	\centering
	\begin{tabular}{|c|c|c|c|c|c|c|} 
		\hline
		Surface & W/O, normal & W/, normal  & W/O, AMD & W/, AMD & \cite{shah2018multiple} normal & \cite{shah2018multiple} AMD  \\ 
		\hline
		IRPE & 2.31$\pm$0.54 & {\bdf 2.21$\pm$0.58} & 3.80$\pm$2.01 & {\bdf 3.63$\pm$1.71} & 3.84$\pm$0.58 & 6.07$\pm$1.84  \\
		\hline
		OBM & 2.97$\pm$0.35 & {\bdf 2.93$\pm$0.37} & 6.09$\pm$3.10 & {\bdf 5.83$\pm$2.89} & 4.97$\pm$1.01 & 5.85$\pm$1.80  \\
		\hline
	\end{tabular}
	\caption {Unsigned mean surface positioning errors (UMSP) for the results on the SD-OCT test dataset. The unit is in $\mu m$. W/O: without SB; W/: with SB.}
	\label{tab:oct}	
	\vspace{-1.cm}
\end{table}
\par
\begin{figure}[htb!]
	\centering
	\includegraphics[width=\textwidth]{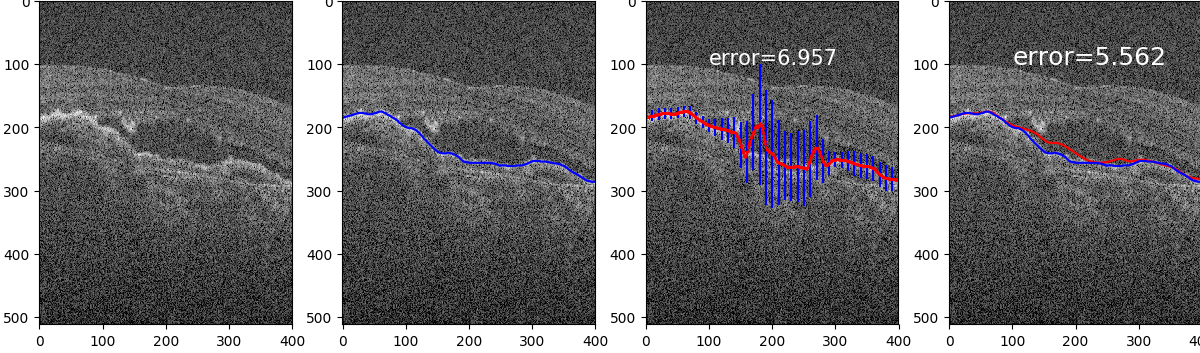}
	\includegraphics[width=\textwidth]{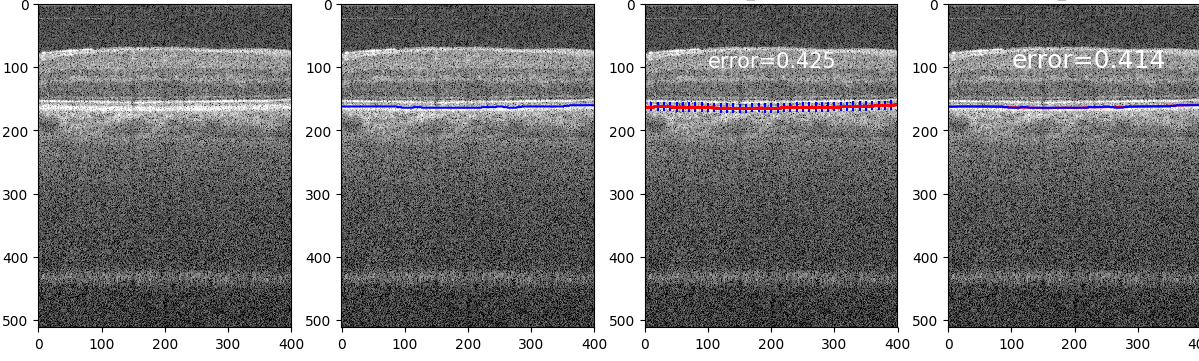}
	\includegraphics[width=\textwidth]{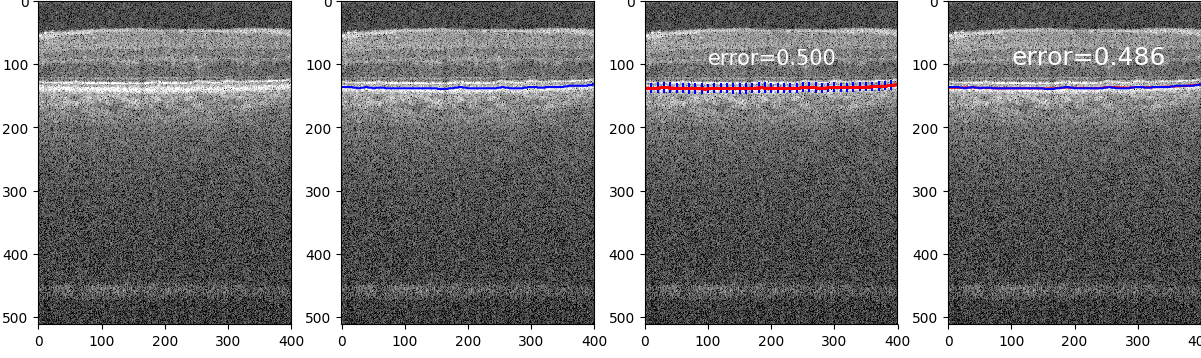}
	\includegraphics[width=\textwidth]{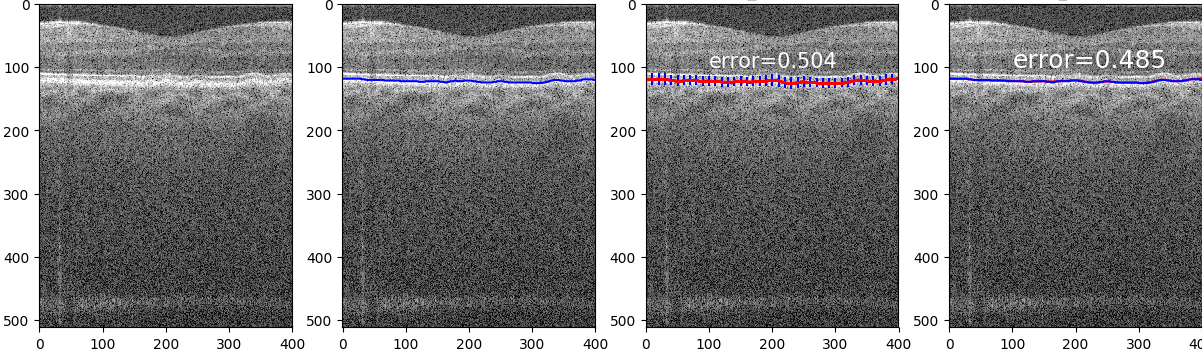}
	\includegraphics[width=\textwidth]{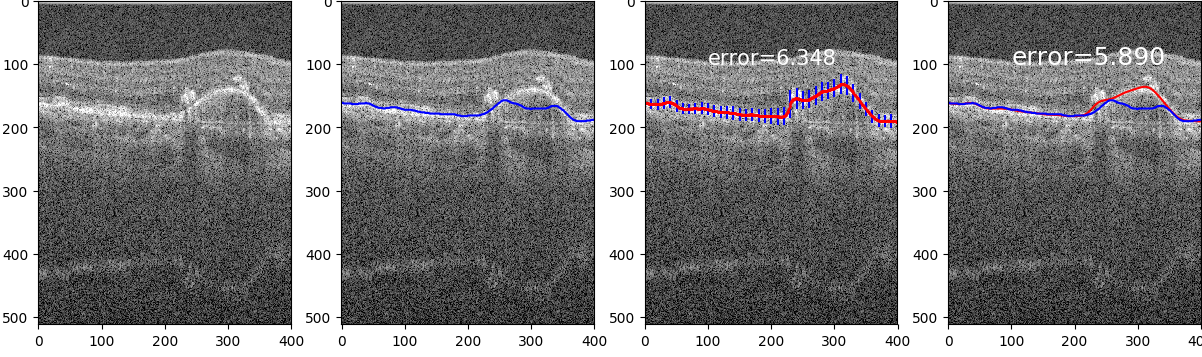}
	\caption{Sample segmentations of the OBM  surface. The four columns illustrate original images, ground truth, predictions without SB (red: Gaussian means, blue: std bars) and that with SB (red: prediction, blue: ground truth). For illustrations, the standard deviation bars are downsampled. The errors shown are in the unit of pixel.}
	\label{fig:obm_seg}
\end{figure}

\begin{figure}[htb!]
	\vspace{-0.5cm}
	\centering
	\includegraphics[width=\textwidth]{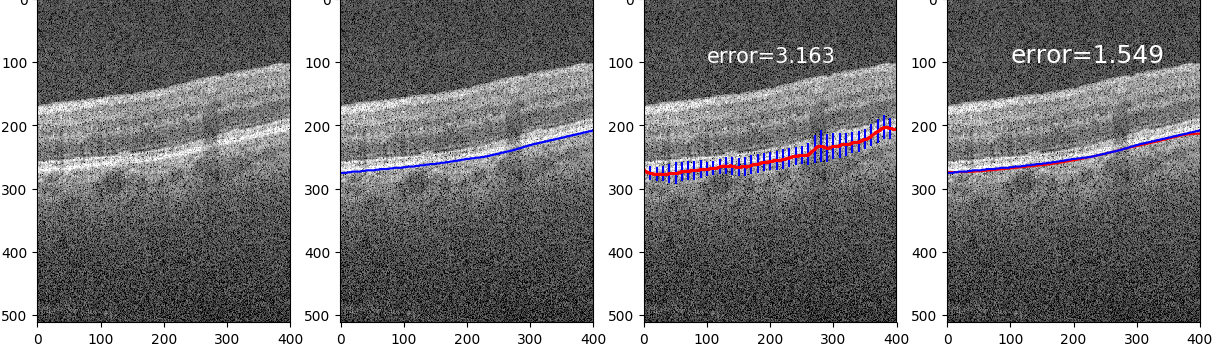}
	\includegraphics[width=\textwidth]{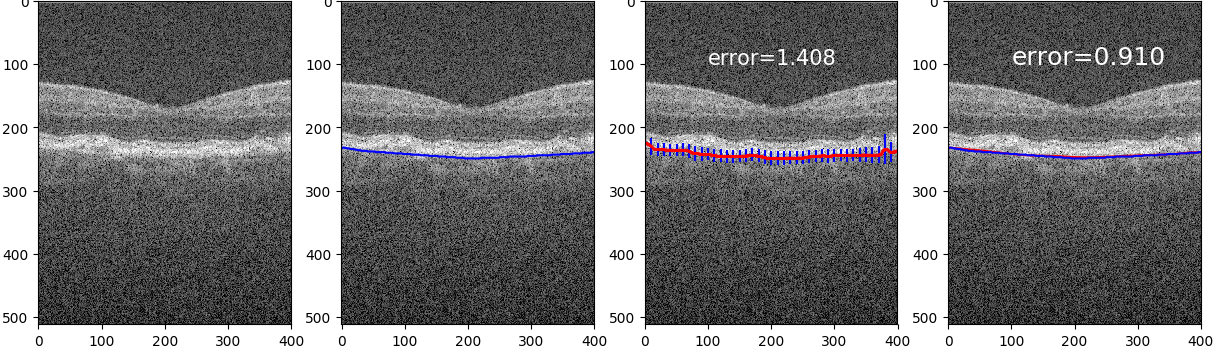}
	\includegraphics[width=\textwidth]{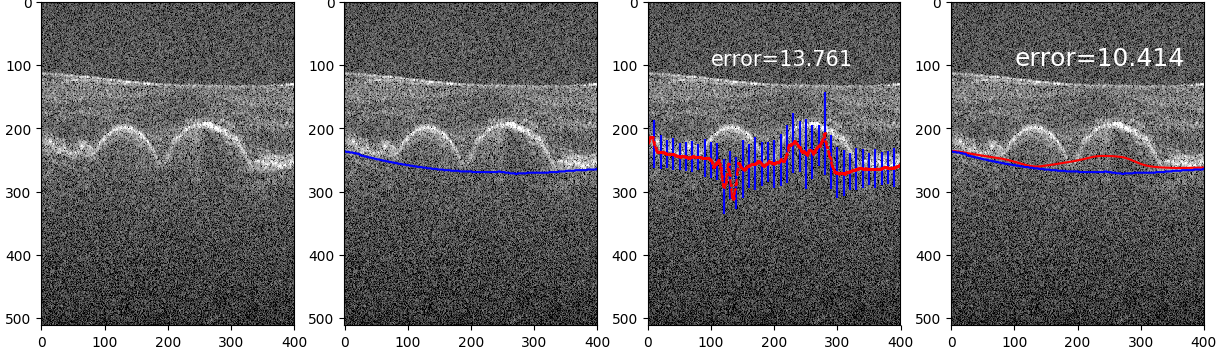}
	\includegraphics[width=\textwidth]{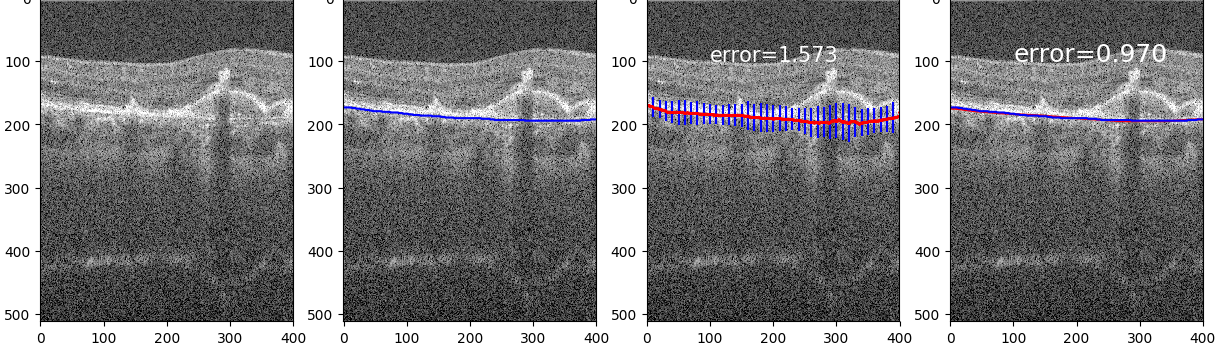}
	\includegraphics[width=\textwidth]{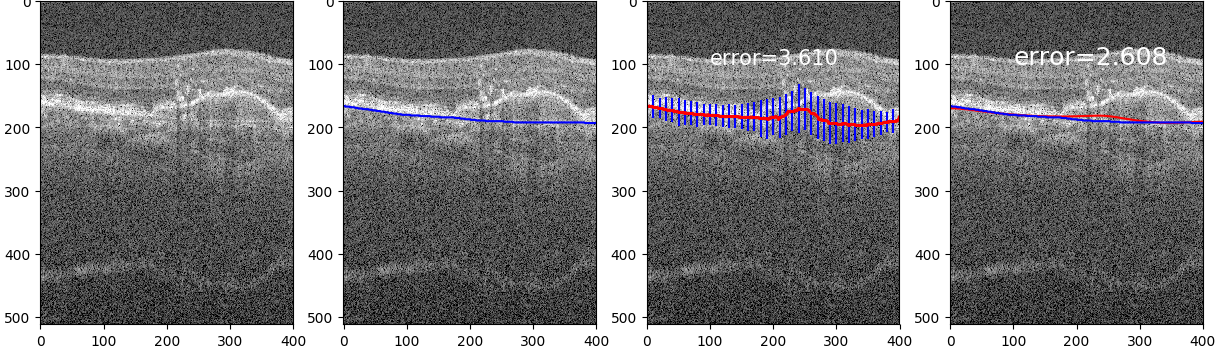}
	\caption{Sample segmentations of the IRPE  surface. The four columns illustrate original images, ground truth, predictions without SB (red: Gaussian means, blue: std bars) and that with SB (red: prediction, blue: ground truth). For illustrations, the standard deviation bars are downsampled. The errors shown are in the unit of pixel.}
		\label{fig:irpe_seg}
	\vspace{-0.5cm}
\end{figure}

\noindent {\bf Results}
Unsigned mean surface positioning error (UMSP)~\cite{garvin2009automated}  is utilized  for evaluation of segmentation accuracy. The quantitative results are summarized in Table~\ref{tab:oct}. We compare to another deep learning based method~\cite{shah2018multiple}, which is the state-of-the-art on this dataset. As for the IRPE surface, the proposed method without SB outperforms~\cite{shah2018multiple}. With SB plugged in, the performance can be improved further: for the normal cases, the UMSP can be improved by 39\%; for the AMD cases, we can achieve a similar improvement (40\%). The segmentation of the OBM surface is more challenging than that of the IRPE surface. One can notice that the learned SB can improve the segmentation accuracy consistently. As for the OBM surface, compared to~\cite{shah2018multiple}, the proposed method achieves 41\% improvement on the normal cases and performs comparably on the AMD cases. An interesting observation is that the learned $w_{comp}$ are  around 0.05 and 1.0 for IRPE and OBM, respectively, which coincides with the human observation that generally less context information exists for OBM.
Sample segmentation results are illustrated in Fig.~\ref{fig:oct}, Fig.~\ref{fig:obm_seg} and Fig.~\ref{fig:irpe_seg}.

\subsection{IVUS Vessel Wall Segmentation}
The data used for this experiment was obtained from the standardized evaluation of IVUS image segmentation database~\cite{balocco2014standardized}. In this experiment, the dataset B was used. This dataset consists  of 435 images with a size of $384\times384$, as well as the respective expert manual tracings  of lumen and  media surfaces. It comprises two groups - a training set (109 slices) and a testing set (326 slices). 
\begin{wrapfigure}{R}{0.32\linewidth}
	\centering
	\vspace{-0.9cm}
	\includegraphics[width=0.32\textwidth]{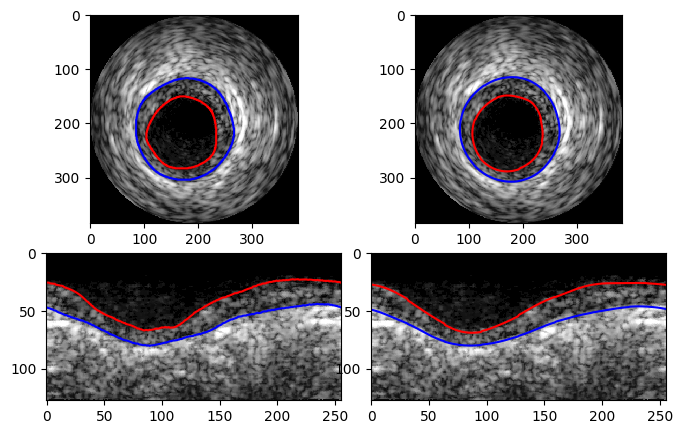}	
	\vspace{-0.7cm}
	\caption{Sample IVUS data. Ground truth (left) and predictions (right) in Cartesian (top) and Polar system (bottom). Red: lumen; Blue: media.}
	\label{fig:polar}
\vspace{-.9cm}
\end{wrapfigure}
The experiment with the proposed method was conducted in conformance with the directives provided for the IVUS challenge. In our experiment, we randomly split the 109 training slices into 100 slices for training and 9 slices for validation. 
\noindent {\bf Pre-processing and Augmentation}
Each slice was transformed to be represented in the polar coordinate system with a size of $256\times 128$, as illustrated in Fig.~\ref{fig:polar}. The intensity of each slice  was normalized to have a zero mean and a unit standard deviation.
The Gaussian truth was generated for each {\em column} using $\sigma=15$, as it has a shorter column (128 vs 512 in the SD-OCT data). 
As the  number of training data was limited, we augmented the data on the fly by random combinations of various operations including mirroring, circulation shifting along the polar dimension, adding Gaussian noises (mean=0, std=0.1), adding Salt and Pepper noises (5\%), and cropping (90\% of the original size) and then resizing ($256\times 128$). All training parameters were the same as those for the SD-OCT segmentations, except that the number of epochs for the pre-training was 4000.
\begin{table}
	\vspace{-.7cm}
	\centering
	\begin{tabular}{|c|c|c|c|c|c|c|} 
		\hline
		Methods & \multicolumn{3}{|c|}{Lumen} &   \multicolumn{3}{|c|}{Media}   \\ 
		\hline
		& JM  & PAD & HD & JM  & PAD & HD  \\
		\hline
		P3\cite{cardinal2006intravascular,cardinal2010fast}  & {\bdf 0.88$\pm$0.05}  & {\bdf 0.06$\pm$0.05} & 0.34$\pm$0.14 & {\bdf 0.91$\pm$0.04}  & {\bdf 0.05$\pm$0.04} & {\bdf 0.31$\pm$0.12} \\
		\hline
		VGG-U-net\cite{balakrishna2018automatic} & 0.80(-)  & - & - & 0.81(-)  & - & - \\
		\hline
		\cite{shah2019optimal} & 0.86$\pm$0.04  & 0.09$\pm$0.03 & 0.37$\pm$0.14 & 0.90$\pm$0.03  & 0.07$\pm$0.03 & 0.43$\pm$0.12 \\
		\hline
		W/O SB  & 0.87$\pm$0.11 & 0.08$\pm$0.12 & 0.43$\pm$1.90 & 0.88$\pm$0.10  & 0.08$\pm$0.10 & 0.41$\pm$0.31\\
		\hline
		W/ SB  & {\bdf 0.88$\pm$0.06} & 0.07$\pm$0.07 & {\bdf 0.28$\pm$0.18} & 0.89$\pm$0.07  & 0.07$\pm$0.07 & 0.37$\pm$0.27\\
		\hline
	\end{tabular}
	\caption {Segmentation results comparison on the IVUS dataset. }
	\label{tab:ivus_test}
	\vspace{-1.2cm}	
\end{table}
\par
\begin{figure}
	\vspace{-.8cm}
	\centering
	\includegraphics[width=\textwidth]{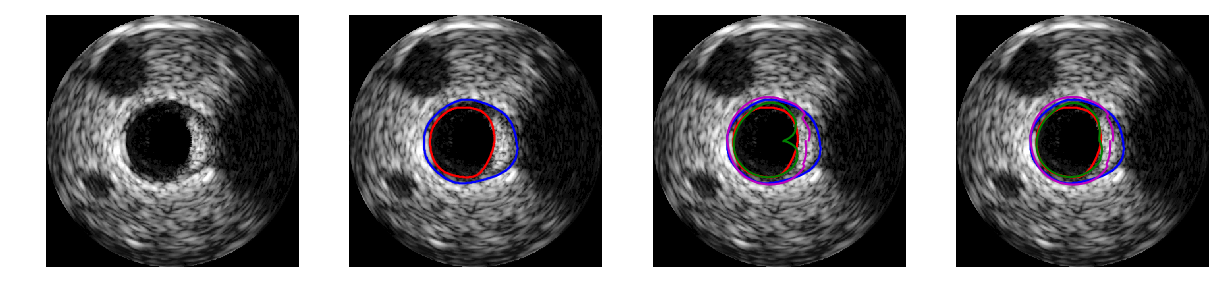}
	\centering
	\includegraphics[width=\textwidth]{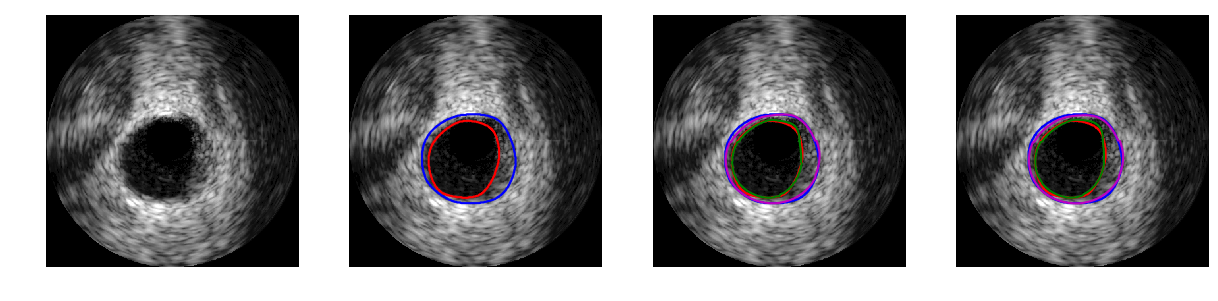}
	\centering
	\includegraphics[width=\textwidth]{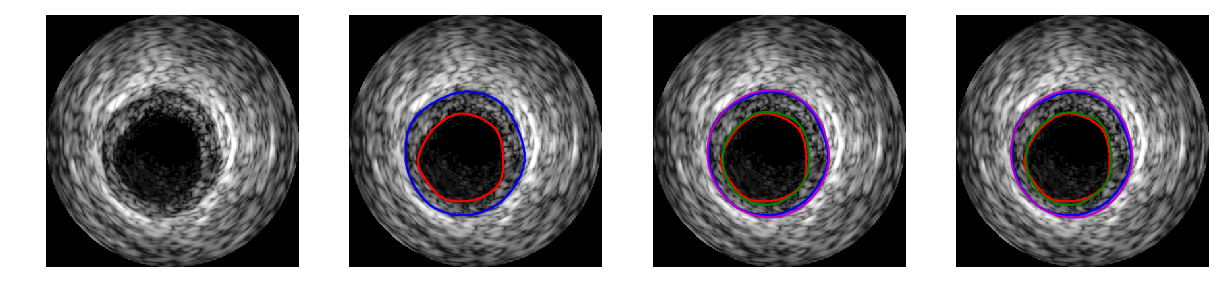}
	\centering
	\includegraphics[width=\textwidth]{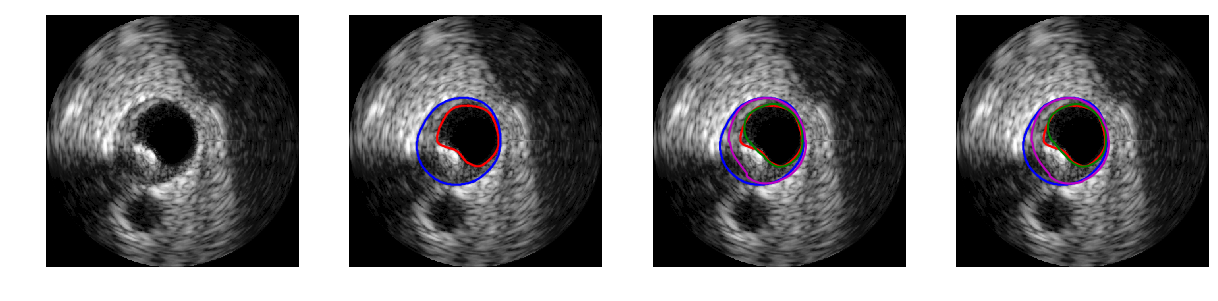}
	\includegraphics[width=\textwidth]{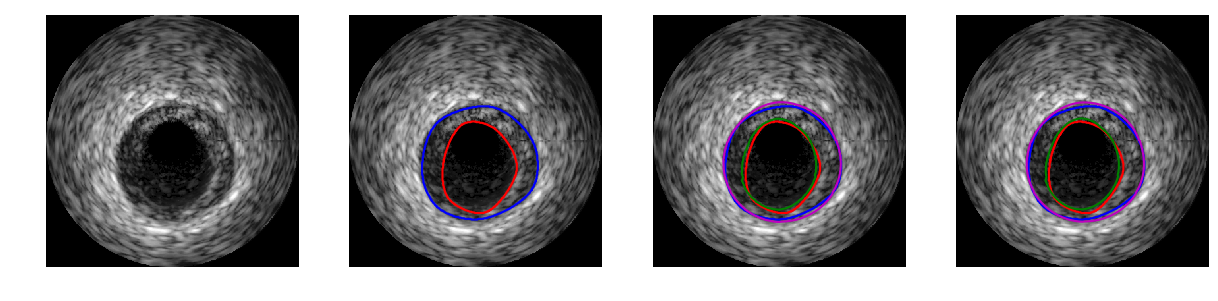}
	\includegraphics[width=\textwidth]{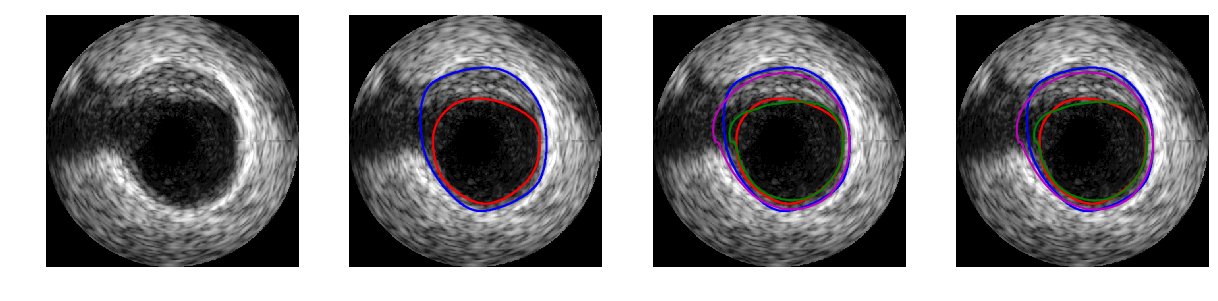}
	~ 
	\caption{Sample segmentation results on IVUS data. Red: lumen ground truth; Blue: media ground truth; Green: lumen prediction; Purple: media prediction. The four columns correspond to the original image, the ground truth, the prediction without SB, the prediction with SB.}
	\label{fig:ivus_seg}
	\vspace{-0.5cm}
\end{figure}

 \noindent {\bf Results}
Jaccard Measure (JM), Percentage of Area Difference (PAD) and Hausdroff Distance (HD) are utilized to evaluate segmentation accuracy. The results are summarized in Table~\ref{tab:ivus_test}. P3~\cite{cardinal2006intravascular,cardinal2010fast} is the state-of-the-art method for this IVUS dataset. It is an expectation maximization based method and is {\em semi automated}. VGG-U-net denotes a deep learning based method \cite{balakrishna2018automatic}. The state-of-the-art fully automation method is a graph search based method working in irregularly sampled space \cite{shah2019optimal}, the unary energy of which is learned with random forests.
From Table~\ref{tab:ivus_test}, one can find that the proposed method outperforms the sub-voxel graph based method~\cite{shah2019optimal}  for the lumen surface and the performance on the media surface is comparable. Compared to the state-of-the-art semi-automate method P3, the performance of the proposed is comparable  for the lumen surface and is marginally inferior on the media surface. The proposed method outperforms the  VGG-U-net by a big margin. Sample segmentation results are illustrated in Fig.~\ref{fig:polar} and Fig.~\ref{fig:ivus_seg}.  
\par 
As to the inference computation time,  the proposed method needs more overhead than the VGG-U-net method (0.21 vs 0.09 sec/slice). The overhead is mainly from the Hessian matrix computation, as well as two separate runs of the program for two surfaces. While compared to P3~ \cite{cardinal2006intravascular,cardinal2010fast} (8.6 sec/slice) and the graph based method~\cite{shah2019optimal} (187.4 sec/slice), the proposed method is highly efficient.
\section{Conclusion}
In this paper, a  novel segmentation model based on a convolutional neural network (CNN) and a learnable surface smoothing block  
is proposed to tackle the surface segmentation problem with end-to-end training. To the best of our knowledge, this is the first study for surface segmentation which can achieve guaranteed globally optimal solutions using deep learning. Experiments  on SD-OCT retinal layer segmentation and IVUS vessel wall segmentation demonstrated very promising results. The proposed method is applicable to  $2$-D and $3$-D.
\bibliography{ref_short}
\end{document}